\documentclass[twocolumn,amsmath,amssymb,pra]{revtex4-1}
\usepackage{amsmath} % need for subequations

    \usepackage{graphicx}
    \usepackage{subfigure}
    \usepackage{color}
    \usepackage{dcolumn}
    \usepackage{bm}
    \usepackage{xspace}
    \usepackage{hyperref}

\newcommand{\SOM}[1]{\textcolor{red}{\cite{SOM}}}

\newcommand{\emphCaption}[1]{{\bf{#1}}}
\begin{document}

\title{Comment on ``Wigner phase space description of a Morse oscillator'' [J. Chem. Phys. {\bf 77}, 4604 (1982)]}

\author{Dimitris Kakofengitis, Maxime Oliva and Ole Steuernagel}

\affiliation{School of Physics, Astronomy and Mathematics, University of Hertfordshire, Hatfield, AL10 9AB, UK 
%\institute{School of Physics, Astronomy and Mathematics, University of Hertfordshire, Hatfield, AL10 9AB, UK
\email{O.Steuernagel@herts.ac.uk}
}

\date{\today}

\maketitle

In reference~\cite{Lee_Scully_JCP82} Lee and Scully introduce the concept of trajectories
for the study of quantum dynamics in quantum phase space. Specifically, they consider
energy eigenstates (of the Morse potential) of the quantum-mechanical Wigner
distribution~\cite{Wigner_32,Hillery_PR84}.

They state that ``\emph{the main purpose of our investigation is to obtain physical
  insights, we consider a rather trivial case when no external force is applied to the
  oscillator. Then, the oscillator should remain in the same eigenstate throughout.  In
  terms of the Wigner distribution, it means that each phase-space point should move in
  such a way that the Wigner distribution does not change in time. This consideration
  leads to the concept of `Wigner trajectories', trajectories along which phase-space
  points of the Wigner distribution move. For the case under consideration, Wigner
  trajectories must be trajectories along the surfaces on which the Wigner distribution
  takes on the same value, i.~e., trajectories along the equi-Wigner surfaces. These
  Wigner trajectories are `quantum-mechanical' trajectories in the sense that they
  represent paths of phase-space points that move according to the quantum-mechanical
  equation of motion. They describe the exact quantum-mechanical dynamics in a phase
  space, whereas classical trajectories obviously yield only an approximate description of
  quantum dynamics}.''\cite{Lee_Scully_JCP82}

This concept of `Wigner trajectories' was referred to by many and explicitly endorsed by
some~\cite{Razavy_PLA96,Razavy_Book_03} but also
criticised~\cite{Daligault_PRA03,Dittrich_Pachon_JCP10}. Specifically, Dittrich \emph{et al.}
consider it meaningless because their semi-classical integration
method~\cite{Dittrich_Pachon_JCP10} did not produce `Wigner trajectories'.  They find
their own semi-classical trajectories, starting from classical trajectories, when
increasingly refined, at first approach the ``\emph{Wigner contour [...]. However, they do
  not approach it asymptotically but continue shifting further past the Wigner contour,
  indicating that it plays no particular role for quantum time evolution in phase space,
  not even of eigenstates.}''\cite{Dittrich_Pachon_JCP10}

This still leaves the question whether Lee and Scully are correct?

Here we show that Lee and Scully's concept of trajectories for energy eigenstates of
anharmonic quantum mechanical systems is flawed~\cite{Oliva_arXiv161103303O}. Instead,
there is a well defined alternative concept: Wigner's phase space current~$\bm
J$~\cite{Ole_PRL13,Kakofengitis14,Kakofengitis_Div16}.

This current can always be integrated and yields fieldlines which resemble the
`trajectories' Lee and Scully tried to find. The $\bm J$-fieldlines show behaviour very different
from what Lee and Scully speculated might happen: the fieldlines neither follow the
contours of the Wigner distribution nor are the values of the Wigner distribution along
the fieldlines constant.

The time evolution of~$W$, Wigner's quantum phase-space
distribution~\cite{Wigner_32,Hillery_PR84}, is given by the Eulerian continuity
equation~\cite{Wigner_32,Oliva_arXiv161103303O} (also called the quantum `Liouville'
equation, although it is not Liouvillian~\cite{Kakofengitis_Div16})
\begin{flalign} 
  \partial_t W(\bm r,t) = - \bm \nabla \cdot {\bm J(\bm r,t)} \; .
\label{eq:W_ContinuityMT}
\end{flalign}
Above, partial derivatives (abbreviated as $\partial_t = {\partial}/{\partial t} )$ are
combined to form the divergence $\bm \nabla \cdot \bm J = \partial_x J_x + \partial_p
J_p$.

Generally, Wigner current~$\bm J$ has an integral
representation~\cite{Wigner_32,Hillery_PR84,Baker_PR58}, but for potentials~$V(x)$ that
can be Taylor-expanded, giving rise to finite forces only, $\bm J$ is of the
form~\cite{Wigner_32}
\begin{eqnarray} {\bm J} = \binom{ J_x }{ J_p } = \bm j +
 \begin{pmatrix} 0
    \\-\sum\limits_{l=1}^{\infty}{\frac{(i\hbar/2)^{2l}}{(2l+1)!}
      \partial_p^{2l}W \partial_x^{2l+1}V }
 \end{pmatrix}
\; .
\label{eq:FlowComponentsMT}
\end{eqnarray}
Here, with $\bm v = \binom{ {p}/{M} }{- \partial_x V }$, $\bm j = W \bm v$ is the
classical term and $\bm J - \bm j$ are quantum terms.

\begin{figure*}[t]
\centering
\includegraphics[width=148mm,angle=0]{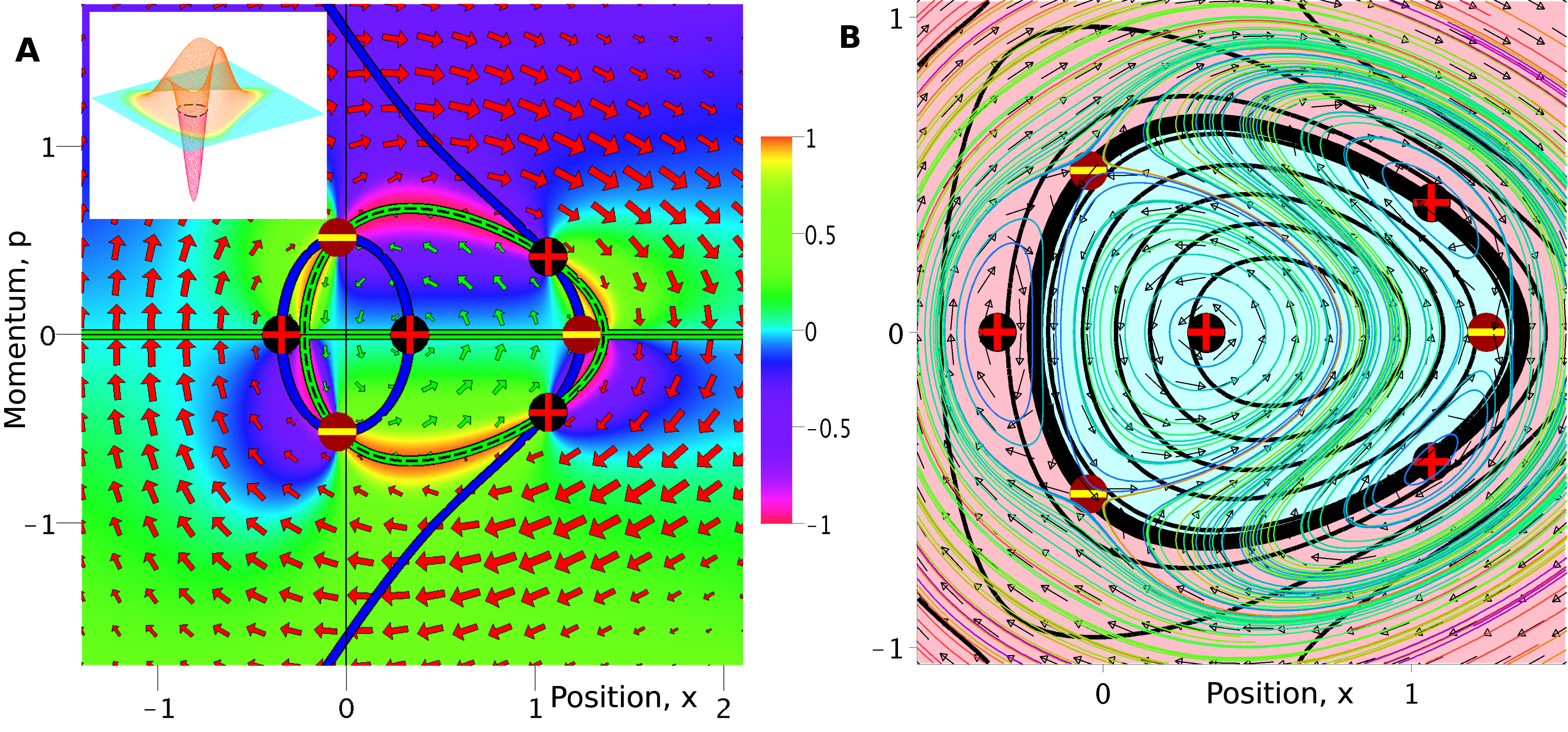}
  \caption{\textsf{\textbf{A}}, \emphCaption{Singularities of~$\bm \nabla \cdot \bm w$
      coincide with Zeros of~$W$.} ${\bm J}$ depicted by arrows (red for clockwise and
    green for inverted flow~\cite{Ole_PRL13}), together with the zeros of the $J_x$ and
    $J_p$ components (green and blue lines, respectively), is superimposed on top of a
    colourplot of $\frac{2}{\pi} \arctan (\bm \nabla \cdot \bm w)$. The inset shows the
    corresponding Wigner distribution for the first excited state of an anharmonic Morse
    oscillator~\cite{Dahl_JCP88} with
    potential~$V(x)=3(1-\exp({{-x}/{\sqrt{6}}}))^2$. The red crosses and yellow bars
    mark the locations of the flow's stagnation points, with Poincar\'e-Hopf
    indices~\cite{Ole_PRL13}~$\omega=+1$ and~$-1$. Parameters:~$\hbar=1$ and $M=1$. The black
    dashed line marks the zero-contour of the Wigner distributions (compare inset), here 
the divergence~$\bm \nabla \cdot \bm w$ is singular~\cite{Oliva_arXiv161103303O}.  
\newline \textsf{\textbf{B}},
    \emphCaption{Integrated Fieldlines of $\bm J$ cross Wigner Distribution
      Contours}. Thin coloured lines display fieldlines of $\bm J$, displayed together
    with normalized current ${\bm J}/||\bm J||$ (black arrows), and its stagnation points, for the
    same state as depicted in~\textsf{\textbf{A}}. $W$'s zero contour, around the
    negative (light cyan-coloured) patch at the centre, is highlighted by a thick black
    line. Many fieldlines, for this first excited state, cut across the Wigner
    distribution's contours and enter and leave the negative area.
    \label{fig:divergence}}
\end{figure*}

Fieldlines of Wigner current are well defined and their depiction has helped to reveal the
topological charge conservation of~$\bm J$'s stagnation points~\cite{Ole_PRL13}, see
Fig.~\ref{fig:divergence}.

The concept of trajectories, instead of $\bm J$-fieldlines, originates from phase space
transport in Lagrangian form using the total (or comoving) derivative~$\frac{d W}{d t}$.
To investigate this transport we have to decompose the continuity
equation~(\ref{eq:W_ContinuityMT}) in Lagrangian
form~\cite{Donoso_PRL01,Trahan_JCP03,Daligault_PRA03,Oliva_arXiv161103303O}
\begin{equation} 
  \frac{d W}{d t} = \partial_t W + \bm w \cdot \bm \nabla W =- W \bm \nabla \cdot \bm w \; . 
\label{eq:W_TotalDerivMT}
\end{equation}
Here, the quantum phase space velocity field~$\bm
w$~\cite{Donoso_PRL01,Trahan_JCP03,Daligault_PRA03}, corresponding to the hamiltonian
velocity field~$\bm v$, is
\begin{equation}
\bm w = \frac{\bm J}{W} =\bm v + \frac{1}{W}
 \begin{pmatrix} 0
    \\-\sum\limits_{l=1}^{\infty}{\frac{(i\hbar/2)^{2l}}{(2l+1)!}
      \partial_p^{2l}W \partial_x^{2l+1}V } 
 \end{pmatrix} \; .
 \label{eq:Wigner_w_velocityMT}
\end{equation}
$\bm w$ and its divergence
\begin{equation}
  \label{eq:WignerFlow_VelocityDivergence}
  \bm{\nabla} \cdot \bm{w} = \frac{W\bm{\nabla} \cdot \bm{J} - \bm{J}
    \cdot \bm{\nabla}W}{W^2} \;  
\end{equation}
are singular at zeros of~$W$, since generally zeros of~$W$ do not coincide with
zeros of its derivatives~\cite{Oliva_arXiv161103303O} 

Lee and Scully's assumption that \emph{``Wigner trajectories must be trajectories along
  the surfaces on which the Wigner distribution takes on the same value''} formally
amounts to the statement that $\bm{w}\cdot\bm{\nabla}W=0$. According to
Eq.~(\ref{eq:Wigner_w_velocityMT}) this implies~ $\bm{J}\cdot\bm{\nabla}W=0$
(if~$W\neq0$).  For eigenstates $ \bm{\nabla}\cdot \bm{J} = - \partial_tW =0 $, the flow
is therefore assumed to be Liouvillian~$ \bm{\nabla} \cdot \bm{w} =0$ (see
Eq.~(\ref{eq:WignerFlow_VelocityDivergence})): \emph{``each phase-space point should move
  in such a way that the Wigner distribution does not change in time''},
(formally~$\frac{dW}{dt} = 0$, see Eq.~(\ref{eq:W_TotalDerivMT})).

Both statements are incorrect, as the Morse oscillator is
anharmonic, generally $\bm{\nabla} \cdot \bm{w} \neq 0$~\cite{Kakofengitis_Div16}
and~$\frac{dW}{dt}\neq0$.

Elsewhere~\cite{Oliva_arXiv161103303O} we have shown that anharmonic quantum mechanical
systems do not feature trajectories because the values of $\bm w$ are, according to
Eq.~(\ref{eq:Wigner_w_velocityMT}), singular when~$W=0$.  Additionally, the divergence of
the velocity field features singularities,
see~Fig.~\ref{fig:divergence}~\textsf{\textbf{A}}, which cannot be transformed
away~\cite{Kakofengitis_Div16}. Here, we confirm those abstract conclusions by a plot of
the fieldlines of~$\bm J$ in Fig.~\ref{fig:divergence}~\textsf{\textbf{B}}. This shows
that Wigner current crosses from positive to negative areas and back. In other words,
unlike in the classical case, even for energy eigenstates~$\frac{dW}{dt}\neq0$.

It might seem as if our analysis on the one hand and Lee and Scully's on the other are
each internally consistent. It is therefore worth explaining where exactly Lee and Scully
went wrong. Arguably, their starting point is Eq.~(3.19) in
reference~\cite{Lee_Scully_JCP82}. In our terminology their Eq.~(3.19) states that the
second component of~$\bm w$ has the form $w_p = \partial_x V_{eff}(x,p) =
\frac{J_p}{\partial_p W}$, this is incorrect, the correct form is given
in Eq.~(\ref{eq:Wigner_w_velocityMT}).

Lee and Scully arrived at this incorrect decomposition of~$\bm J$ to yield their version
of~$w_p$ because they have assumed that quantum phase space flow is Liouvillian, i.e., has
the form~$\partial_t W = -\frac{p}{M} \partial_x W + \partial_x V_{eff} \partial_p W $
(their Eq.~(3.13)). But this is inconsistent, since constructing~$\bm J = W \bm w$ from
their version of~$w_p$ does \emph{not} yield Eq.~(\ref{eq:W_ContinuityMT}) as the
evolution equation.

Lee and Scully's decomposition was criticised by Daligault~\cite{Daligault_PRA03},
criticised and yet adopted by Sala \emph{et al.}~\cite{Sala_JCP93} and by Henriksen
\emph{et al.}~\cite{Henriksen_CPL88} (who later concluded though that, based on numerical
work, \emph{``These studies showed a fatal degradation of the distribution
  function''}~\cite{Moller_JPC94}). Their decomposition was also adopted by, e.g., Muga
\emph{et al.}~\cite{Muga_SSC95}, Razavy~\cite{Razavy_PLA96,Razavy_Book_03}, Dias and
Prata~\cite{Dias_JMP02}, Zhang and Zheng~\cite{Zhang_ChPL09}, and reported by
Landauer~\cite{Landauer_RMP94}.

Confusion persists about the nature of quantum phase space dynamics: confusion about the
correct decomposition of the continuity equation, the fact that trajectories do not
exist~\cite{Oliva_arXiv161103303O}, and that the quantum Liouville
equation~(\ref{eq:W_ContinuityMT}) in Lagrangian decomposition~(\ref{eq:W_TotalDerivMT})
features singular divergence of its velocity field~$\bm w$ (which cannot be
removed~\cite{Kakofengitis_Div16}).

Wigner current~$\bm J$ and its fieldlines are a powerful tool to study the behaviour of
quantum phase space dynamics. Particularly, plots like those in Fig.~\ref{fig:divergence}
and elsewhere~\cite{Ole_PRL13,Kakofengitis14} have guided our understanding of quantum
phase space dynamics --- to share this observation is our main motivation for this
comment.

%\bibliography{/home/ole/Shared/Open/AA/META/bibliographies/WignerFlow}

\end{document}